\documentclass[aps,prl,twocolumn,showpacs,superscriptaddress]{revtex4}
\usepackage{graphicx}
\usepackage{color}

\begin{document}

\title{Tailoring the electronic texture of a topological insulator via its surface orientation}
\author{Lucas Barreto}
\affiliation{Department of Physics and Astronomy, Interdisciplinary Nanoscience Center (iNANO), Aarhus University,
8000 Aarhus C, Denmark}
\author{Wendell Simoes e Silva}
\affiliation{Departamento de F\'{\i}sica, Universidade Federal de Minas Gerais, 30123-970, Belo Horizonte, Brazil}
\affiliation{Department of Physics and Astronomy, Interdisciplinary Nanoscience Center (iNANO), Aarhus University,
8000 Aarhus C, Denmark}
\author{Malthe Stensgaard}
\author{S\o ren Ulstrup}
\author{Xie-Gang Zhu}
\author{Marco Bianchi}
\affiliation{Department of Physics and Astronomy, Interdisciplinary Nanoscience Center (iNANO), Aarhus University,
8000 Aarhus C, Denmark}
\author{Maciej Dendzik}
\affiliation{Centre for Nanometer-Scale Science and Advanced Materials (NANOSAM),
Smoluchowski Institute of Physics, Jagiellonian University, Reymonta
4, 30-059 Krak\'ow, Poland}
\affiliation{Department of Physics and Astronomy, Interdisciplinary Nanoscience Center (iNANO), Aarhus University,
8000 Aarhus C, Denmark}
\author{Philip Hofmann}
\affiliation{Department of Physics and Astronomy, Interdisciplinary Nanoscience Center (iNANO), Aarhus University,
8000 Aarhus C, Denmark}
\affiliation{email: philip@phys.au.dk}
\date{\today}
\begin{abstract}
%%\mr{The author list and order are still up for discussion. Put Wendell's affiliation in Brazil?}
Three dimensional topological insulator crystals consist of an insulating bulk enclosed by metallic surfaces, and detailed theoretical predictions about the surface state band topology and spin texture are available \cite{Fu:2007b,Fu:2007c,Moore:2007,Murakami:2007,Teo:2008,Dai:2008,Zhang:2009}. While several topological insulator materials  are currently known, the existence and topology of these metallic states have only ever been probed for one particular surface orientation of a given material. For most topological insulators, such as Bi$_{1-x}$Sb$_{x}$ and  Bi$_2$Se$_3$, this surface is the closed-packed (111) surface and it supports one topologically guaranteed surface state Dirac cone.  Here we experimentally realise a non closed-packed  surface of a topological insulator,  Bi$_{1-x}$Sb$_{x}$(110),  and probe the surface state topology by angle-resolved photoemission. As expected, this surface also supports metallic states but the change in surface orientation drastically modifies the band topology, leading to three Dirac cones instead of one,  in excellent agreement with the theoretical predictions \cite{Teo:2008} but in contrast to any other experimentally studied TI surface. This illustrates the possibility to tailor the basic  topological properties of the surface  via its crystallographic direction. Here it introduces a valley degree of freedom not previously achieved for topological insulator systems.
\end{abstract}

\maketitle

 For the topological insulator (TI) surfaces studied so far, the predicted Fermi contour consists of an odd number of Dirac cones around the Brillouin zone centre $\bar{\Gamma}$ (see e.g. Refs. \cite{Hsieh:2008,Xia:2009,Chen:2009,Kuroda:2012,Neupane:2012}) but more complex scenarios with Dirac cones around three translation invariant momenta (TRIMs) have been predicted for non-(111) surfaces of Bi$_{1-x}$Sb$_{x}$ \cite{Teo:2008} and for SmB$_6$(001) \cite{Lu:2012b}. The possibility of such a scenario is closely linked to the bulk electronic structure of a TI: If the bulk parity inversion only takes place at the Brillouin zone centre $\Gamma$, this leads to an odd number of  closed Fermi contours around the surface Brillouin zone centre $\bar{\Gamma}$ for every possible surface orientation, but if the bulk parity inversion happens at another bulk TRIM, one should be able to change the surface state topology by choosing different surface orientations. Candidates for TI materials allowing this are Bi$_{1-x}$Sb$_{x}$ (parity inversion at every bulk TRIM except L \cite{Teo:2008}), PbBi$_2$Te$_4$ (parity inversion at Z \cite{Kuroda:2012}) and SbB$_6$ (parity inversion at X \cite{Lu:2012b}). 

So far, surface state topology  investigations by angle-resolved photoemission spectroscopy (ARPES) have only been performed for surfaces with a topologically predicted single Dirac cone around $\bar{\Gamma}$. Indeed, only one surface orientation for a given TI has so far been probed, such that even the basic expectation of a bulk material completely enclosed by metallic surfaces still needs to be verified. Experiments for other surface orientations are lacking due to the difficulty of preparing such surfaces. Feasible methods for preparing TI surfaces are cleaving \cite{Hsieh:2008,Xia:2009}, a combination of cutting, polishing and \emph{in situ} cleaning, as well as epitaxial growth \cite{Zhang:2009a,Hirahara:2010,Brune:2011}. While cleaving and epitaxial growth have so far only been used to prepare the very stable (111) surface (except for the cubic HgTe(001)  \cite{,Brune:2011}), cutting and \emph{in situ} cleaning could in principle give any desired surface direction, but the approach is hindered by the small size of crystals, their strong anisotropy and the high likelihood to off-set the delicate stoichiometry near the surface during the  \emph{in situ} cleaning process. 

\begin{figure*}
 \includegraphics[width=.99\textwidth]{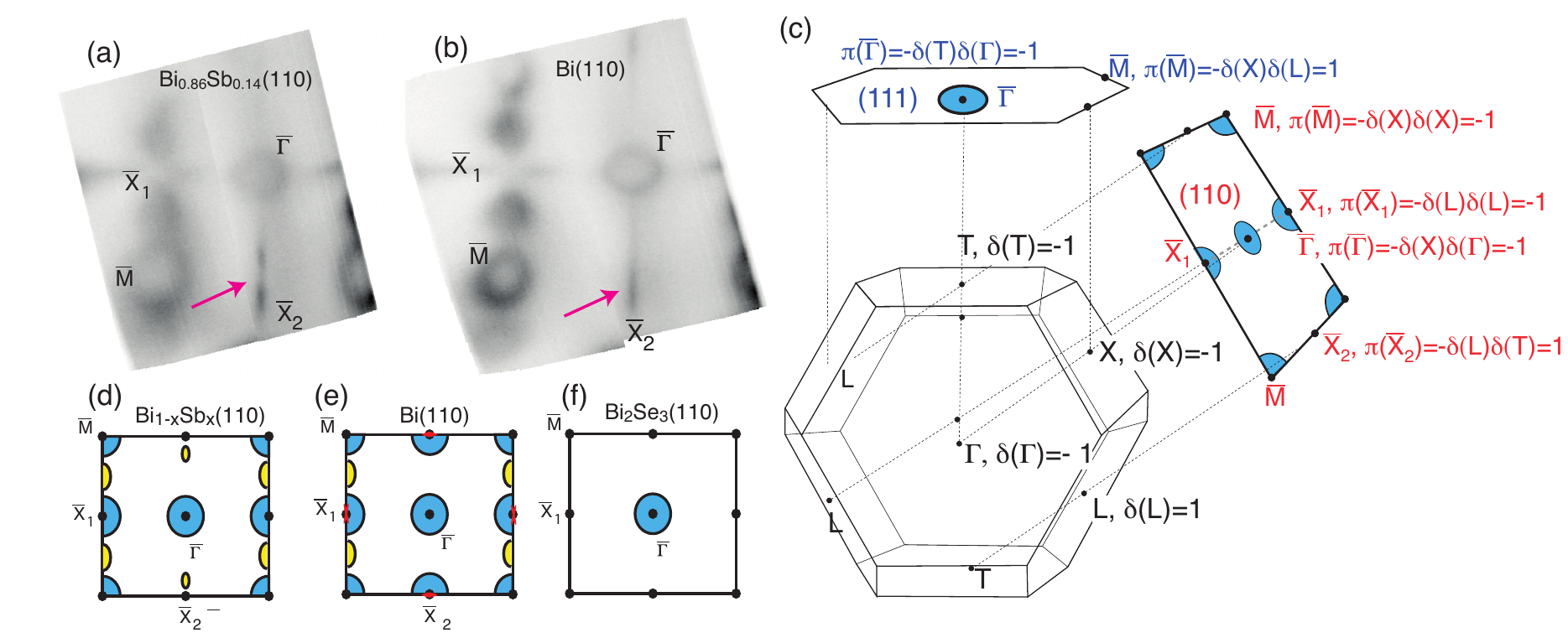}\\
\caption{(a) and (b) Photoemission intensity at the Fermi energy for Bi$_{0.86}$Sb$_{0.14}$(110) and Bi(110), respectively. Dark corresponds to high intensity. The spectroscopic signature of the topological quantum  phase transition is expected close to the $\bar{X}_2$ point where a splitting of a Fermi surface feature is observed (marked by arrows). (c) Projection of the bulk parity invariants $\delta$  onto the (111) and (110) surfaces of Bi$_{1-x}$Sb$_{x}$, resulting in the surface fermion parity values $\pi$ for the time-reversal invariant momenta (TRIMs). The black dots mark the positions of the time-reversal invariant momenta (TRIMs) in the surface Brillouin zone. The blue areas surrounding TRIMs denote an odd number of closed Fermi contours around the TRIM. (d) - (f) Topological prediction for the surface electronic structure of the (110) surfaces of Bi$_{1-x}$Sb$_{x}$ ($x>0.09$), Bi and Bi$_2$Se$_3$, respectively. The yellow contours are additional but topologically insignificant Fermi contour features found in the experiment. The small red features in (e) correspond to the projected bulk Fermi surface. }
  \label{fig:1}
\end{figure*}

Here we show that an epitaxial film of the topological insulator  Bi$_{1-x}$Sb$_{x}$ can be grown in the (110) orientation by evaporating a mixture of Bi and Sb on a Bi(110) surface. Since $x$ is quite small in the TI phase ($0.09 < x < 0.18$), the lattice mismatch between the Bi substrate and the film is so small \cite{Berger:1982} that the growth is quasi-homoepitaxial, resulting in a high-quality Bi$_{1-x}$Sb$_{x}$(110) film. Fig. \ref{fig:1}(a) shows the Fermi contour for a film of Bi$_{0.86}$Sb$_{0.14}$(110) measured by ARPES (i.e. the photoemission intensity at the Fermi level) and it is immediately obvious that for this surface three TRIMs are enclosed by Fermi contours: $\bar{\Gamma}, \bar{M}$ and $\bar{X}_1$, in sharp contrast to the (111) surface of this TI where only the $\bar{\Gamma}$ point is enclosed by an odd number of Fermi contours. 

Before discussing the experimental results in more detail, we briefly review the theoretical predictions for the Bi$_{1-x}$Sb$_{x}$ surfaces by Teo, Fu and Kane \cite{Teo:2008}. Their approach and its results are illustrated in Fig. \ref{fig:1}(c)-(e). The parity invariants $\delta$ for the eight bulk time-reversal invariant momenta (TRIMs) $\Gamma_{i}$ in Bi$_{1-x}$Sb$_{x}$ are all found to be $-1$, except for the bulk L point where $\delta(L)=1$ for Bi$_{1-x}$Sb$_{x}$ with ($ x > 0.09$) and $\delta(L)=-1$ otherwise. The surface fermion parity $\pi$  for each surface TRIM determines the topology of the surface Fermi contour. It  can be calculated from a projection of the bulk parity invariants  using the relation $\pi(\Lambda_a)=(-1)^{n_b}\delta(\Gamma_{i})\delta(\Gamma_{j})$, where $n_b$ is the number of occupied, spin-degenerate bulk bands, i.e. 5 for Bi$_{1-x}$Sb$_{x}$. This procedure is illustrated in  Fig. \ref{fig:1}(c) together with the bulk and surface Brillouin zones. The predicted surface fermion parity values for the surface TRIMs of the (111) and (110) surfaces in the TI state ($ x > 0.09$) are also shown. The surface fermion parity values for the surface TRIMs immediately give the topological prediction for the Fermi contour: the TRIM should be enclosed by an odd number of contours for $\pi=-1$ and by an even or no contour for $\pi=1$. The topologically expected closed contours are marked by blue circles. Bi$_{1-x}$Sb$_{x}$(111) shows one such contour in the surface Brillouin zone whereas Bi$_{1-x}$Sb$_{x}$(110) has three. 
%Note that the situation for Bi$_{1-x}$Sb$_{x}$(111) is the same as for the closed packed surface of e.g. Bi$_2$Se$_3$ despite of the entirely different bulk topology. This is merely a coincidence and related to the fact that Bi$_2$Se$_3$ has an even number of spin-degenerate bulk bands $n_b=14$ whereas Bi$_{1-x}$Sb$_{x}$ has $n_b=5$, leading to a sign change in all the surface fermion parity values.

The topological predictions for Bi$_{1-x}$Sb$_{x}$(110) and Bi(110) are compared in Fig.  \ref{fig:1}(d) and (e).  The quantum phase transition to a TI at $ x = 0.09$ changes the bulk  Fermion parity for the $L$ point from $-1$ to $1$ and this leads to a surface fermion parity change at $\bar{X}_2$: For Bi$_{1-x}$Sb$_{x}$(110)  one expects to find an odd number of Fermi level crossings between $\bar{X}_2$ and any other TRIM whereas there would be an even number for Bi(110). Note  that such a rich surface state topology would not be expected for a non-(111) surface of most other topological insulators (e.g. Bi$_2$Se$_3$) because there the bulk parity inversion happens only at the $\Gamma$ point ($\delta(\Gamma)=-1$), which, combined with the even $n_b$, leads to merely an odd number of closed contours around the $\bar{\Gamma}$ point of the surface, as illustrated for Bi$_2$Se$_3$(110) in Fig. \ref{fig:1}(f). 

The measured Fermi surfaces for Bi(110) and Bi$_{0.86}$Sb$_{0.14}$(110) are shown in Fig. \ref{fig:1}(a) and (b), respectively. They are, at first glance, very similar to each other and to published results for Bi(110) \cite{Agergaard:2001,Pascual:2004}. The dominating features are topologically required closed contours around the $\bar{\Gamma}$, $\bar{M}$ and $\bar{X}_1$ points as well as a small electron pocket on the $\bar{X}_1-\bar{M}$ line. This electron pocket does not encircle any TRIM and is irrelevant for the surface state topology, but in order to facilitate the comparison between the topological predictions and the data, it has been included as a yellow feature in Fig. \ref{fig:1}(d),(e).  The topological prediction for Bi(110) also requires the existence of a closed Fermi contour around $\bar{X}_2$, but this is not clearly confirmed by the data in this overview scan. In any event, such predictions do not necessarily have to be fulfilled for a semimetal because the topologically required parity changes between surface TRIMs could also be accomplished by the bulk states crossing the Fermi surface. Such projected bulk Fermi surface elements are found near the $\bar{X}_1$ and $\bar{X}_2$ points on Bi(110). They are indicated by the red areas in Fig. \ref{fig:1}(e). 

\begin{figure}
\includegraphics[width=.4\textwidth]{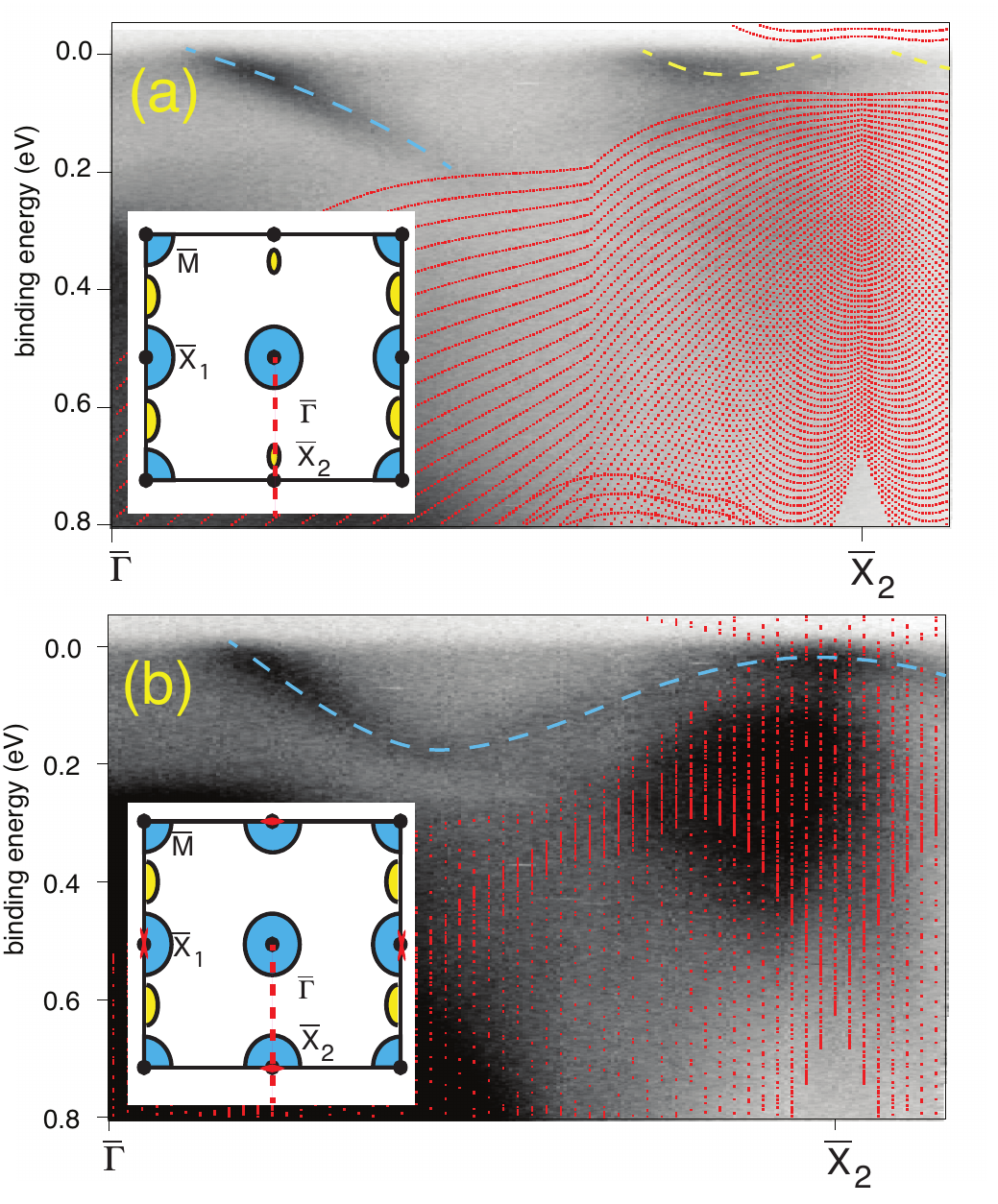}\\
  \caption{Detailed electronic structure along the $\bar{\Gamma}-\bar{X}_2$ line for (a) Bi$_{0.86}$Sb$_{0.14}$(110) and (b) Bi(110). Dispersing surface states are marked by dashed coloured lines as a guide to the eye. The red dots denote the projected bulk band structure. The inset shows the surface Brillouin zone with the topological predictions in blue and additional experimentally found Fermi contours in yellow. The magenta lines in the inset mark the scan direction. }
  \label{fig:2}
\end{figure}

The topological quantum phase transition from Bi to Bi$_{1-x}$Sb$_{x}$ should be reflected in a change of the surface state dispersion around the $\bar{X}_2$ point. This is consistent with the observed change between the Fermi surface maps of Fig. \ref{fig:1}(a) and (b): While the photoemission intensity at the Fermi level near $\bar{X}_2$ is an elongated stripe in the $\bar{\Gamma}-\bar{X}_2$ direction for Bi(110), this stripe is split into two features on either side of $\bar{X}_2$ for Bi$_{0.86}$Sb$_{0.14}$. This situation near $\bar{X}_2$ is further explored by a high-resolution scan along $\bar{\Gamma}-\bar{X}_2$ for both  Bi$_{0.86}$Sb$_{0.14}$(110) and Bi(110) in Fig. \ref{fig:2}. The surface state dispersion in this direction is quite different in the two cases: Both surfaces share the hole pocket around $\bar{\Gamma}$ but whereas the band forming this pocket disappears into the projected band states for  Bi$_{0.86}$Sb$_{0.14}$(110), its dispersion reaches a binding energy maximum for Bi(110) before it disperses back towards the Fermi level. It only merges with the bulk bands in the immediate vicinity of the $\bar{X}_2$ point. For Bi$_{0.86}$Sb$_{0.14}$(110), on the other hand, there is no state observable at $\bar{X}_2$ but a shallow electron pocket very close to this point. 

These differences in the band structure bring the expected topology change about. For Bi(110), one would expect the $\bar{X}_2$ point to be encircled by an odd number of Fermi contours. Experimentally, the situation is somewhat unclear because it cannot be determined if the surface state band barely crosses the Fermi level on both sides of $\bar{X}_2$, forming a small hole pocket around the point, or if it stays occupied. In any case, this is not important because one would not necessarily expect the topological predictions to be strictly fulfilled to for a semimetal surface. Any required change of parity could be brought about by the bulk band continuum near $\bar{X}_2$. For Bi$_{0.86}$Sb$_{0.14}$(110), on the other hand, the topological predictions should be strictly followed. In contrast to the situation for Bi(110), $\bar{X}_2$ should not be encircled by a closed Fermi contour and it clearly is not: the shallow electron pocket marked by a yellow line is close to the $\bar{X}_2$ point but it does not encircle it. Indeed, the electron pockets on either side of $\bar{X}_2$ can clearly be seen in the Fermi surface scan of Fig. \ref{fig:1}(a). 

% compare to the situation on Bi(111)?

\begin{figure}
  \includegraphics[width=.4\textwidth]{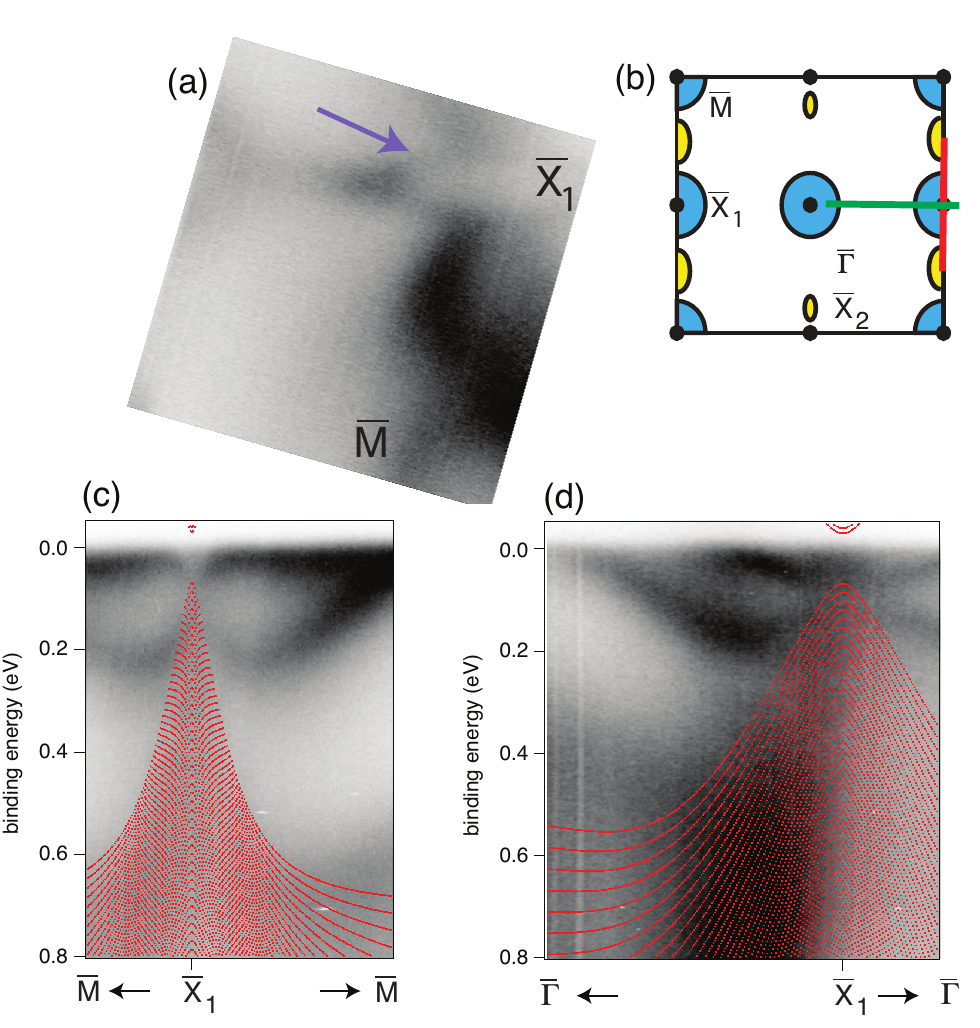}\\
  \caption{Detailed electronic structure around the $\bar{X}_1$ point of Bi$_{0.86}$Sb$_{0.14}$(110), confirming the existence of a closed Fermi contour around this point. (a) Photoemission intensity at the Fermi level. The arrow marks the closed contour. (b) Sketch of the surface Brillouin zone with the scan directions marked by a green and a magenta line. For the colour coding of the schematically indicated Fermi surface features, see caption of Fig. \ref{fig:1}. (c) Band structure around $\bar{X}_1$ towards the $\bar{M}$ points (magenta line in (b)). (d) Band structure around $\bar{X}_1$ towards the $\bar{\Gamma}$ points (green line in (b)). The closed Fermi contour around $\bar{X}_1$ is found to be an electron pocket. A Dirac point is formed at $\bar{X}_1$, but it lies inside the projected continuum of the bulk valence band. The projected band structure is outlined by the red dots.}
  \label{fig:3}
\end{figure}

A final topological prediction for Bi$_{0.86}$Sb$_{0.14}$ is the existence of a closed Fermi contour around $\bar{X}_1$. This can already be seen in Fig. \ref{fig:1}(a) but is shown more clearly by a detailed inspection of the band structure around this TRIM in Fig. \ref{fig:3}. The high-resolution Fermi contour in Fig. \ref{fig:3}(a) already shows a closed contour around $\bar{X}_1$ in the immediate vicinity of the two electron pockets along $\bar{X}_{1}-\bar{M}$. Scans around $\bar{X}_1$ towards $\bar{M}$ and $\bar{\Gamma}$ reveal that this feature is another electron pocket. In fact, the band giving rise to it merges with the lower Rashba-split surface state bands originating from $\bar{M}$ and $\bar{\Gamma}$ to form a Dirac point at $\bar{X}_1$ but this Dirac point is situated in the projected valence band. This situation and the shape of the Fermi contour around $\bar{X}_1$ is similar to the situation on Sb(110), i.e. the case of $x=1$ \cite{Bianchi:2012}.

%The surface state topology with three Dirac cones is also confirmed by a simple model calculation, using a transfer-matrix approach and the same interpolation of the tight-binding parameters as in Ref. \cite{Teo:2008}. The result of this calculation is given in Fig. \ref{fig:4}. The same three TRIMs as in the experimental data are found to be encircled by closed Fermi contours: $\bar{\Gamma}$,  $\bar{M}$ and $\bar{X}_1$. However, the detailed dispersion is quite different from the experimental results with the Dirac points at $\bar{\Gamma}$ and $\bar{M}$ below the Fermi energy in the calculation and above in the experiment. Indeed, the agreement is notably worse than for the case of  Bi$_{0.86}$Sb$_{0.14}$(111) \cite{Hsieh:2008,Teo:2008}. This is not unexpected because the formation of the (111) surface does not require the breaking of any pseudo-covalent bonds within the bilayers forming the Bi structure, whereas this is needed for (110). This bond breaking can be expected to lead to a more severe deviation of the hopping parameters near the surface from those that adequately describe the bulk. However, this inferior agreement does not affect the surface state topology. 
%
%\begin{figure}
%  % Requires \usepackage{graphicx}
%  \includegraphics[width=.75\textwidth]{fig4}\\
%  \caption{Simple model calculation of the electronic structure of Bi$_{0.86}$Sb$_{0.14}$(110) along lines connecting the TRIMs.}
%  \label{fig:4}
%\end{figure}

The entire Fermi contour of Bi$_{0.86}$Sb$_{0.14}$(110) is thus in excellent agreement with the theoretical predictions for this surface. It is also substantially more complex than that of  Bi$_{1-x}$Sb$_{x}$(111) or the predictions of the non-(111) surfaces of Bi$_2$Se$_3$-type topological insulators. Indeed, this illustrates the design possibilities when exploiting the surface orientation of a topological insulator. The example shown here is particularly interesting because it adds a valley degree of freedom to the topological Dirac fermions, something that is currently attracting great attention in the case of graphene \cite{Rycerz:2007}. It should also be possible to use the same strategy for the construction of quasi one-dimensional surface states on three-dimensional topological insulators, as already found for the vicinal parent surface Bi(114) \cite{Wells:2009}.

\section{Methods}
Experiments were performed at the SGM-3 beamline of the synchrotron radiation source ASTRID \cite{Hoffmann:2004}. 
Thin films of Bi$_{1-x}$Sb$_{x}$(110) were grown \emph{in situ} on a Bi(110) substrate by simultaneous evaporation of Bi and Sb from an e-beam source. The Bi(110) surface had previously been cleaned by cycles of Ne ion sputtering and annealing. Data were taken for different film thicknesses and compositions. The data shown here were obtained from a film with a thickness of approximately 25~double layers and $x=0.14$ (here the term double layer is used to group the first and second layer of the A7(110) surface. These are almost at the same height \cite{Jona:1967,Hofmann:2006,Sun:2006}). The growth rate and chemical composition of the film were determined by inspecting the shallow Sb~4d and Bi~5d core levels, as in Ref. \onlinecite{Hirahara:2010}. The energy and angular resolution for ARPES measurements were better than 20 meV and 0.2$^{\circ}$, respectively. The photon energy was 28.5~eV for the data shown in Fig. \ref{fig:1} and 18~eV otherwise. Data were collected for many different photon energies between 14~eV and 32~eV in order to make sure that the features discussed above as surface states are in fact surface states, i.e. that their dispersion does not depend on the photon energy. The sample temperature during the measurements was 80~K. The bulk Fermi surface projections and band structure surface projection were calculated using the tight-binding parameters of Liu and Allen \cite{Liu:1995}. For the alloy, the parameters were interpolated following the method in Ref. \onlinecite{Teo:2008}. The calculation of the surface band structure follows the transfer matrix method used in Ref. \onlinecite{Teo:2008} but employs a Green's function method to calculate the spectral function for a semi-infinite crystal \cite{Mele:1978}.

\section{Acknowledgements}
We gratefully acknowledge financial support from the Carlsberg foundation,  the VILLUM foundation, The Danish Council for
Independent Research / Technology and Production Sciences. WSS acknowledges support by the  the Brazilian funding agencies CNPq and CAPES.

%
%\bibliographystyle{phunsrt}
%\bibliography{groupreferences_new}

\end{document}